\documentclass[preprint]{revtex4}

\begin{document}

\title{Spacetime quantization induced by axial currents}

\author{S. Carneiro$^1$ and M. C. Nemes$^2$}

\affiliation{$^1$Instituto de F\'{\i}sica, Universidade Federal da
Bahia, 40210-340, Salvador, BA, Brazil \\ $^2$Departamento de
F\'{\i}sica, ICEX, Universidade Federal de Minas Gerais, CP702,
30123-970, Belo Horizonte, MG, Brazil}

\begin{abstract}
In the present contribution we show that the introduction of a
conserved axial current in electrodynamics can explain the
quantization of electric charge, inducing at the same time a
dynamical quantization of spacetime.
\end{abstract}

\maketitle

\section{Introduction}

In 1931 Dirac proposes an electromagnetic theory with magnetic
monopoles \cite{Dirac}, whose appeal is mainly connected to the
possibility of explaining the quantization of electric charge. In
spite of this undeniable theoretical appeal, in Dirac's theory one
is faced with a symmetry problem: the terms responsible for the
magnetic monopole in the generalized Maxwell equations violate
their symmetry under space and time reversal.

In this work we investigate the introduction of a new conserved
current, namely an axial current, which presents the following
differences as compared to the previously proposed vector magnetic
current: (i) the resulting theory preserves space and time
inversion invariance; (ii) besides the charge quantization, we can
obtain a dynamical quantization of spacetime.

\section{The axial current}

We start with the generalized definition of the electromagnetic
field tensor
\begin{equation}
F_{\mu\nu}=\partial_{\mu}A_{\nu}-\partial_{\nu}A_{\mu}
+\epsilon_{\mu\nu\alpha\beta}\partial^{\alpha}B^{\beta},
\end{equation}
where $B^{\mu}$ represents a new gauge field \cite{Cabibbo}.
Maxwell's equations for the fields $A^{\mu}$ and $B^{\mu}$ in
Lorenz's gauge $(\partial^{\mu}A_{\mu}=\partial^{\mu}B_{\mu}=0)$
then reads
\begin{eqnarray}
\partial^{\nu}F_{\mu\nu}= -\partial^{\nu}\partial_{\nu}A_{\mu}=j_{\mu},\\
\partial^{\nu}F_{\mu\nu}^{\dagger}= -\partial^{\nu}\partial_{\nu}B_{\mu}=g_{\mu},
\end{eqnarray}
where $F_{\mu\nu}^{\dagger}$ corresponds to $F_{\mu\nu}$'s dual
tensor.

The quantity $F_{\mu\nu}$ in $(1)$ is a tensor;
$\epsilon_{\mu\nu\alpha\beta}$ is a pseudo-tensor, and therefore
the field $B_{\mu}$ must be a pseudo-vector or an axial field.
From the point of view of quantum theory the field $B_{\mu}$
represents photon-like particles except for $P$, $T$ and $C$
parities. In other words, axial photons. From this it follows that
$(3)$ is not invariant under time and space reversal, unless
$g^{\mu}$ is also a pseudo-vector \cite{Mignaco}. This suggests
the introduction of an axial current given by
\begin{equation}
g_{\mu}=g\bar{\psi}\gamma_{5}\gamma_{\mu}\psi,
\end{equation}
where $\psi$ represents a spin $1/2$ particle (an axial monopole)
with axial charge $g$.

Since $F_{\mu\nu}^{\dagger}$ is antisymmetric one gets from $(3)$
\begin{equation}
\partial^\mu g_\mu =0,
\end{equation}
which means axial current conservation and therefore massless
axial monopoles.

\section{Charge quantization}

Let us consider the gauge-invariant wave function of a charged
particle \cite{Cabibbo} moving in the presence of the axial
monopole's field,
\begin{eqnarray}
\Phi_{e}\left(x,P'\right)  =  \Phi_{e}\left(x,P\right)
\exp\left[-\frac{\imath e}{2}\int_{S} F^{\mu\nu}d\sigma_{\mu\nu}
\right],
\end{eqnarray}
$S$ being any surface with contour $P'-P$.

Due to the arbitrariness of the surface $S$ we can write
\begin{equation}
\Phi_{e}\left(x,P\right) \exp\left[-\frac{\imath e}{2}\int_{S}
F^{\mu\nu}d\sigma_{\mu\nu} \right]= \Phi_{e}\left(x,P\right)
\exp\left[-\frac{\imath e}{2}\int_{S'} F^{\mu\nu}d\sigma_{\mu\nu}
\right],
\end{equation}
which leads to the condition
\begin{equation}
\exp\left[-\frac{\imath e}{2}\oint_{S-S'}
F^{\mu\nu}d\sigma_{\mu\nu} \right]=1,
\end{equation}
or equivalently to
\begin{equation}
\exp\left[-\imath e\int_{V}
\partial^{\nu}F_{\mu\nu}^{\dagger}dV^{\mu} \right]=1,
\end{equation}
where $V$ is the volume involved by the closed surface $S-S'$.

Using $(3)$, we have
\begin{equation}
\exp\left[-\imath e\int_{V}g_{\mu}dV^{\mu}\right]=1,
\end{equation}
which gives
\begin{equation}
Q_V \equiv \int_V g_{\mu} dV^{\mu} = \frac{2 \pi n}{e},
\end{equation}
$n$ being any integer. Then, using our definition of $g_{\mu}$,
eq. $(4)$, we get
\begin{equation}
Q_{V}=\int_{V}\left(g\bar{\psi}\gamma_{5}\gamma_{\mu}\psi\right)
dV^{\mu}.
\end{equation}

As $Q_{V}$ is a Lorentz scalar, we can perform the calculation in
a convenient reference frame. Taking the axial monopole's frame
(we can do that formally, even the axial monopole being massless)
and using the standard representation for Dirac's spinor,
\begin{equation}
\psi = \left(
\begin{array}{c}
\phi\\ \chi
\end{array}
\right)= \left(
\begin{array}{c}
\phi\\ 0
\end{array}
\right),
\end{equation}
we obtain
\begin{equation}
Q_V = g \int_V (\phi^{\dagger}\;\sigma_i\;\phi)\;dV^i,
\end{equation}
where $\sigma_i$ corresponds to the $i^{th}$ Pauli matrix.

Taking now the axial monopole polarization axis in the direction
of charge's velocity (positive z-axis, say) we get
\begin{equation}
Q_V = g \int \phi^{\dagger}\;\phi\;dx\;dy\;dt.
\end{equation}

Since the axial monopole is massless, the charge's velocity
relative to it has to be necessarily $1$, the velocity of light.
Thus $dt=dz$, and $(15)$ leads to
\begin{equation}
Q_V = g \int \phi^{\dagger}\;\phi\;dx\;dy\;dz= g.
\end{equation}

Equations $(11)$ and $(16)$ give
\begin{equation}
\frac{eg}{2\pi} = n.
\end{equation}
Note that this condition does not depend on the distance between
the electric charge and the axial monopole. It implies in charge
quantization, in the same way of Dirac's charge quantization
condition \cite{Dirac}.

\section{The mass generation}

As we have shown (see $(5)$), in the present context the axial
monopole is necessarily massless, opposed to the massive solitonic
descriptions of magnetic monopoles
\cite{Polyakov,tHooft,Gross,Sorkin}. Let us see what happens if we
circumvent such a restriction.

We shall do that through a dynamical mass generation mechanism, by
introducing a Higgs scalar field with non-vanishing vacuum
expectation value. The Yukawa coupling between the Higgs field and
the axial monopole will generate a mass term for the latter, but
preserving the axial current conservation.

The free lagrangean for the massless axial monopole is
\cite{lagrangiana}
\begin{equation}
{\cal L}_0 = i \bar{\psi} \partial_{\mu} \gamma^{\mu} \psi.
\end{equation}
This lagrangean is invariant under the $U_A(1)$ transformations
defined by
\begin{equation}
U_A(1):\; \psi \rightarrow e^{i\alpha\gamma_5} \psi,
\end{equation}
and this invariance leads to the axial current conservation, eq.
$(5)$.

Now, we add to this free lagrangean the Higgs and Yukawa terms, to
obtain
\begin{equation}
{\cal L} = {\cal L}_0 + {\cal L}_{Higgs} - G \bar{\psi}_L \phi_H
\psi_R - G \bar{\psi}_R \phi_H^{\dagger} \psi_L,
\end{equation}
where $\phi_H$ stands for the Higgs scalar field, $\psi_L$ and
$\psi_R$ are, respectively, the left and right component of
$\psi$, and $G$ is the Yukawa coupling constant.

The lagrangean $(20)$ leads to a massive Dirac equation for the
axial monopole wave function $\psi$, with a mass term given by
\begin{equation}
M=GV,
\end{equation}
with $V$ standing for the vacuum expectation value of the Higgs
field.

It is easy to see that ${\cal L}$ is invariant under the $U_A(1)$
transformations $(19)$, provided they transform the Higgs field as
\begin{equation}
\phi_H \rightarrow e^{-i\alpha} \phi_H.
\end{equation}
Using Noether's theorem we can obtain the conserved current
associated to this invariance. It is precisely our axial current
$(4)$.

\section{Dynamical quantization of spacetime}

Let us investigate the consequences of the mass generation on the
electric charge quantization condition. If the axial monopole is
not massless, the charge velocity relative to it, $v$, is not
necessarily $1$, and now we have $dt=dz/v$. If we consider an
impact parameter sufficiently large, the charge velocity remains
unaffected, and from $(15)$ we obtain
\begin{equation} \label{QV}
Q_V = \frac{g}{v} \int \phi^{\dagger}\;\phi\;dx\;dy\;dz=
\frac{g}{v}.
\end{equation}

Inserting $(23)$ in $(11)$ we have, rather than $(17)$, the
condition
\begin{equation}
\frac{eg}{2\pi v} = n,
\end{equation}
that, again, is independent on the distance between the electric
charge and the axial monopole.

The above relation can be satisfied if we simultaneously fulfill
\begin{equation}
\frac{eg}{2\pi} = n_0,
\end{equation}
and
\begin{equation}
v=\frac{n_0}{n},
\end{equation}
with
\begin{equation}
n\;=\;n_0,\;n_0+1,\;n_0+2...,
\end{equation}
$n_0$ being an integer fixed by the values of $e$ and $g$.

Equation $(25)$ is the charge quantization condition $(17)$,
already derived in the massless case. It can be formally obtained
from $(24)$ if we consider the limit in which the mass of the
particle carrying electric charge vanishes. Or, in another way, if
we ``switch off" the axial monopole mass, taking the false Higgs
vacuum, in which $V=0$. Physically we do not expect charge
quantization to depend on the mass of the particles or on any mass
generation mechanism. We shall therefore assume $(25)-(27)$ as the
only physical solution of $(24)$.

Condition $(26)$ restricts the values of charge's velocity to
rational numbers, a result integrated in theories of discrete
spacetime \cite{Horzela}. Furthermore, these rational values form
a discrete sequence given by $(27)$. For sufficiently high $n_0$,
this sequence tends to a continuum, except for velocities very
near $1$, the light's velocity.

If we consider a massive charged particle, equations $(26)$ and
$(27)$ lead to an upper limit for the particle's velocity, given
by
\begin{equation}
v_0=\frac{n_0}{n_0+1}<1,
\end{equation}
since for a massive particle it is impossible to have $v=1$. For
$n_0\gg1$, this limit is very near $1$.

This limitation of $v$ leads to upper limits for $p$ and $E$, the
momentum and energy of such a particle. For $n_0\gg1$ these upper
limits are
\begin{equation}
p_0\approx E_0\approx m\sqrt{\frac{n_0}{2}},
\end{equation}
which are proportional to the particle mass.

The limitation of the energy-momentum space of the particle leads,
through the uncertainty principle, to the quantization of its
spacetime, with a fundamental length given by
\begin{equation}
a \sim \frac{1}{p_0} \approx \frac{\sqrt{2/n_0}}{m}.
\end{equation}

The quantization of spacetime here has a dynamical nature, opposed
to usual theories of discrete spacetime
\cite{Yukawa,Lee,Connes,Madore,Naschie} where it is purely
kinematical in origin. Here a fundamental length arises owing to
the interaction between the charged particle and the axial
monopole. In this aspect, it is akin rather to the pioneer work of
Wataghin \cite{Wataghin} and to others, more recent, results
\cite{Ahluwalia,Doplicher,Sidharth}. Furthermore, the larger the
charge mass the smaller the fundamental length $a$, such that in
the classical limit ($m\gg0$) spacetime will tend to a continuum.

The experimental upper limit $a<10^{-16}$ cm for the fundamental
length of spacetime gives a lower limit for $n_0$. Inserting that
limit in $(30)$, and using for $m$ the electron mass, we get
$n_0>10^{10}$. Now, from $(25)$ and using for $e$ the electron
charge, we obtain a very high lower limit for the axial charge,
$g>10^{12}$.

\section{Conclusion}

The introduction of the axial current opens up several theoretical
perspectives. Well known symmetries in nature, such as space and
time reversal, are preserved, and a new symmetry, namely the
$U_A(1)$ symmetry, arises in the context of electrodynamics.
Charge quantization can be obtained and, in the massive case, it
is shown to be intimately connected to a dynamical quantization of
spacetime.

But what can we say about the observation of such an entity?
Firstly we can say nothing definite about its mass, since the
values of $G$ and $V$ in $(21)$ are unknown. However, what is
known is that producing massive pairs with large opposite coupling
constants is a very difficult experimental task. Moreover, the
coupling of the axial monopole to the electromagnetic field is not
of vector character, but axial (see $(3)$ and $(4)$). This means
in particular that its production and detection would be directly
associated with polarization conditions, which would render its
observation nontrivial, even in the massless case.

\end{document}